\documentclass[superscriptaddress,showkeys,showpacs]{revtex4}
\usepackage{graphicx}
\usepackage[tbtags]{amsmath}

\begin{document}
\title{Constraints on gluon polarization in the nucleon at NLO accuracy.}
\thanks{Partially supported by CONICET, Fundaci\'on Antorchas, UBACYT and 
ANPCyT, Argentina.}
\author{G. A. Navarro} 
\email{gabin@df.uba.ar}
\affiliation{Departamento de F\'{\i}sica,
Universidad de Buenos Aires\\ Ciudad Universitaria, Pab.1 (1428)
Buenos Aires, Argentina}
\author{R. Sassot}
\email{sassot@df.uba.ar}
\affiliation{Departamento de F\'{\i}sica,
Universidad de Buenos Aires\\ Ciudad Universitaria, Pab.1 (1428)
Buenos Aires, Argentina}
\date{{\bf \today}}

\begin{abstract}
We compare constraints on the gluon polarization in the nucleon 
obtained in next to leading order global QCD fits to polarized deep inelastic 
scattering data with those coming from observables more directly linked to 
the gluon polarization, such as the double spin asymmetry measured by 
{\sc Phenix} at RHIC, and high-$p_T$ hadron production studied by COMPASS 
at CERN.   
\end{abstract}


\maketitle

\section{Introduction}
The extent to which gluons are polarized in the nucleon, and consequently the
origin of the nucleon spin, has persisted as an elusive question for almost 
two decades in spite of strenuous experimental efforts and theoretical 
activity \cite{reviews}. 
Although the spin dependent gluon density in principle can be sized in 
inclusive deep inelastic scattering measurements, mainly through the scale 
dependence of the measured asymmetries, this dependence is rather mild 
in the kinematical range accessed by experiments, and conclusions about it 
are also veiled by our ignorance regarding the polarization of the other 
partonic species, which also contribute to the scale dependence, specially 
that of sea quarks. Therefore, even in the most ambitious scenario, inclusive 
deep inelastic scattering data can at most suggest mild constraints on the 
gluon polarization.

In a recent article \cite{deFlorian:2005mw}, we have shown that the enduring
efforts to measure less inclusive observables in deep inelastic scattering 
have finally begun to yield, allowing combined next to leading order global 
QCD fits
to inclusive and semi-inclusive deep inelastic scattering data where sea quarks
and gluons are much more definitely constrained. In the mean time, independent 
measurements of other less inclusive observables, such as pion production in 
polarized proton-proton collisions \cite{panic}, and high transverse momentum 
hadron pair production in deep inelastic scattering have begun to provide more 
direct assessments of the gluon polarization with competing precision 
\cite{Ageev:2005pq,Procureur:2006sg}. It is therefore of great interest to compare the gluon 
polarization estimates coming from both the global analysis of deep inelastic 
scattering data and from the more direct measurements. In the following we 
perform such comparison and 
we find that although preliminary direct measurements still have a moderate
impact in the fits, there is a remarkable agreement and complementarity 
between both approaches, what encourage us to incorporate the forthcoming data 
in future global analysis.

\section{Comparison}

In the case of inclusive and semi-inclusive polarized deep inelastic scattering
(DIS), the next to leading order (NLO) QCD framework  required  to compute the 
respective observables have been available for some time 
\cite{inclusiva,deFlorian:1995fd}, and indeed recent global analyses have demonstrated both the relevance of 
these corrections and also the non negligible impact of most recent 
semi-inclusive data \cite{deFlorian:2005mw, deFlorian:2000bm}. Specifically, 
in reference \cite{deFlorian:2005mw} it has been found that the best global 
QCD fits to combined DIS data constrain the gluon polarization to be 
moderately positive with a first moment of this density $\delta g \equiv \int
dx \,\Delta g$ of $0.680$ at $10\,GeV^2$ with an uncertainty range given 
by $[0.452,0.771]$ for a one-unit increase in $\chi^2$ and by 
$[-0.107,0.807]$ allowing a more conservative $2\%$ variation of $\chi^2$.
In these constraints, both the requirement of positivity of the polarized 
parton densities relative to a modern set of unpolarized parton densities 
\cite{MRST02}, and the correlation between gluon and sea quark polarization,
 are found to be crucial. Fits with a wide variation in the gluon 
polarization reproduce inclusive data equally well, however they are clearly 
differentiated because of their sea quark polarization by semi-inclusive data.
    
\begin{figure}[hbt]
\includegraphics[width=3.5in]{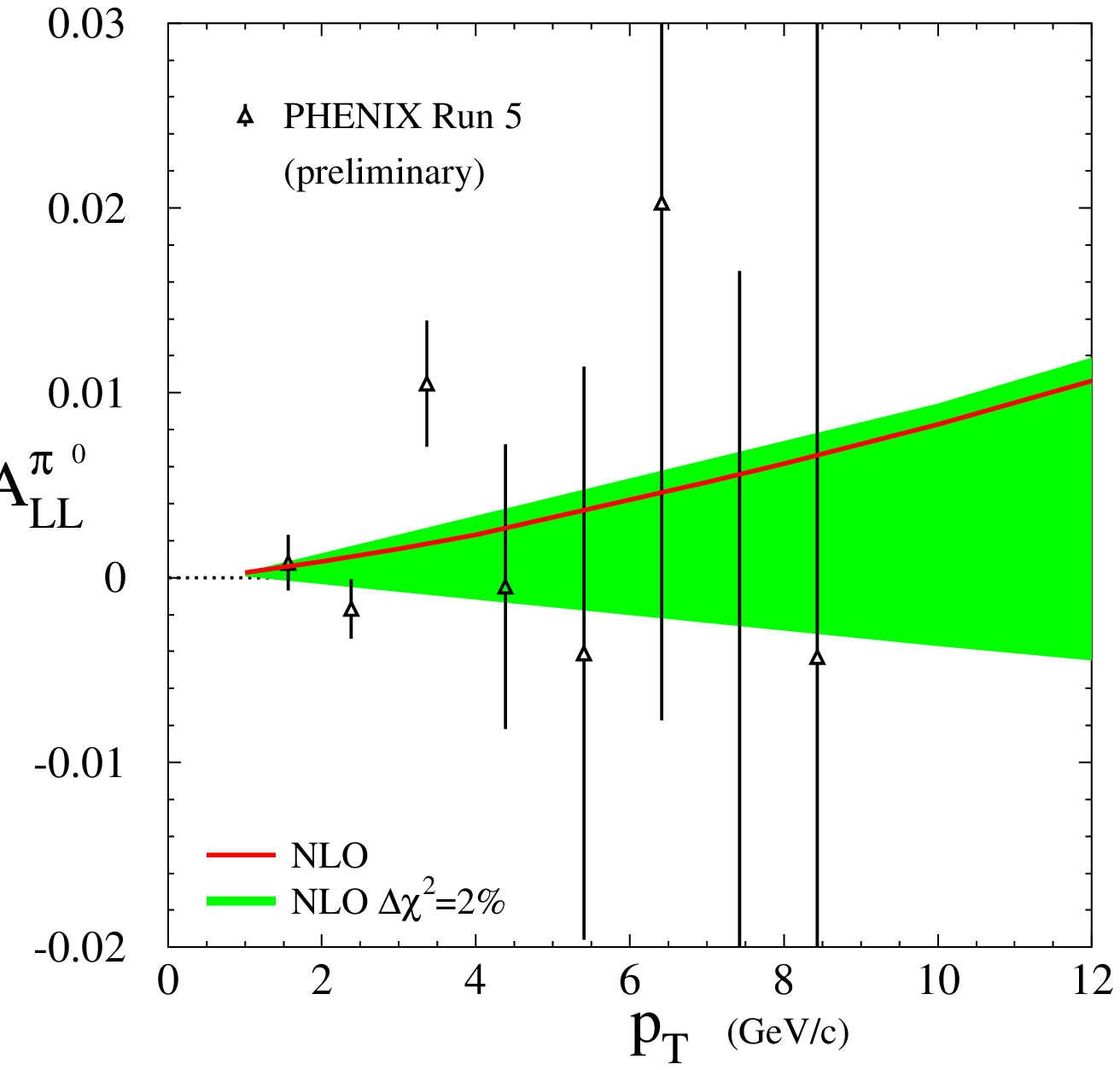}
\includegraphics[width=3.5in]{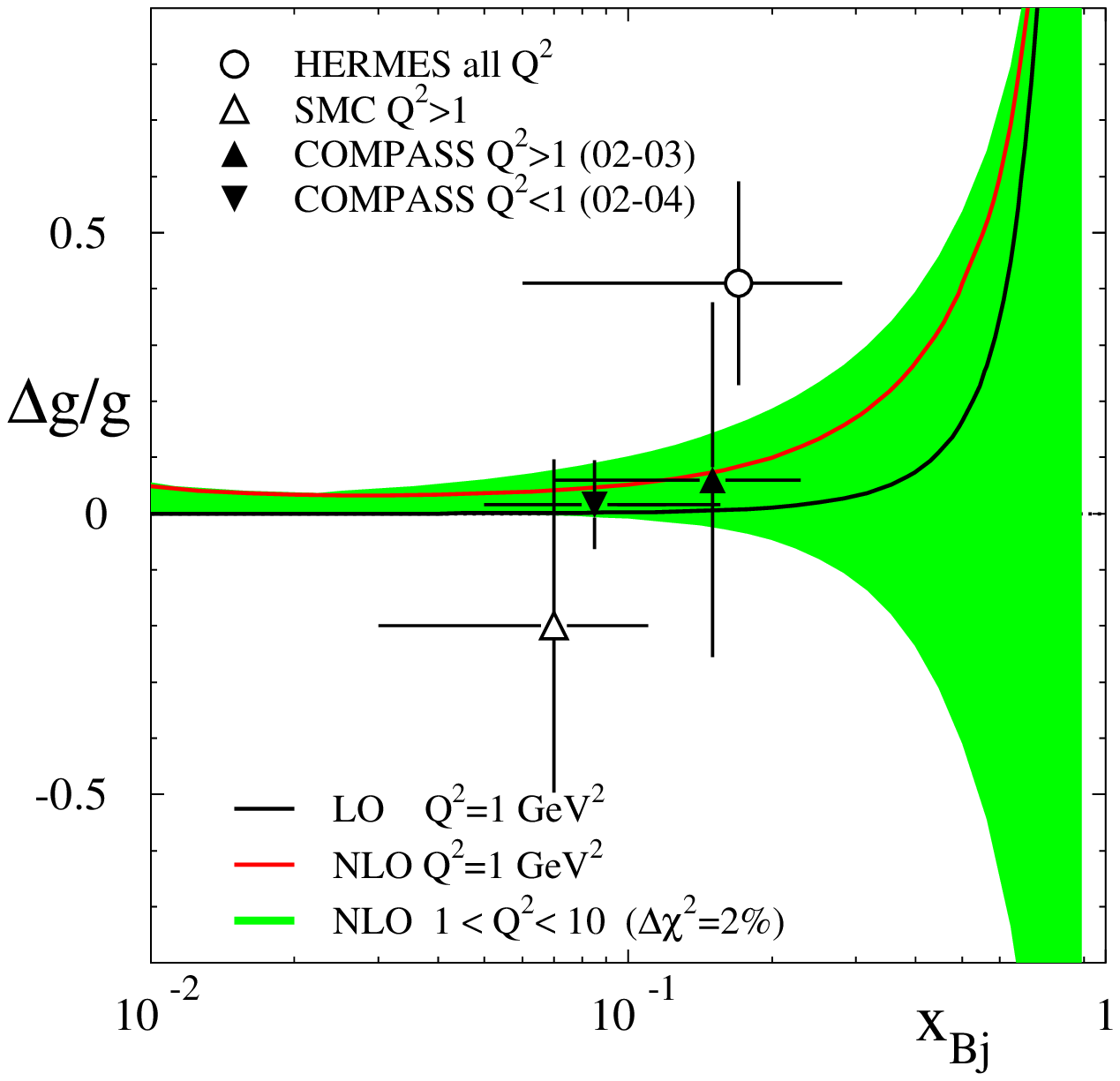}
\caption{ (a) Data on $A^{\pi^0}_{LL}$ \cite{panic} compared with the estimate 
coming from the the NLO fit of reference \cite{deFlorian:2005mw} and the uncertainty band allowing $\Delta \chi^2= 2\%$; (b) the same but for estimates for $\Delta g/g$ from \cite{Procureur:2006sg} .
\label{deltag}}
\end{figure}

The cross section for single inclusive large $p_T$ pion production in 
longitudinally polarized proton-proton collisions, which is right now being 
measured at Brookhaven National Laboratory Relativistiv Heavy Ion Collider 
(BNL RHIC) \cite{Adler:2004ps} have also been computed at NLO 
accuracy, and have been found to be significantly dependent on $\Delta g$ 
\cite{Jager:2002xm,deFlorian:2002az}.
Recently, the {\sc Phenix} collaboration has presented preliminary results with considerably 
reduced errors \cite{panic} which clearly disfavors scenarios with large gluon 
polarization and are in nice agreement with estimates of updated polarized 
fits. In Figure~\ref{deltag}a  we show the expectation for the
double spin asymmetry computed with the best fit of reference 
\cite{deFlorian:2005mw}, together with the data reported by {\sc Phenix} \cite{panic}. 
We also plot the uncertainty band associated to a $\Delta \chi^2= 2\%$ 
variation.
\setlength{\unitlength}{1.mm}
The $\chi^2$ for {\sc Phenix} data obtained with the best fit of reference 
\cite{deFlorian:2005mw}, which is previous to the latest set of {\sc Phenix}
 data, 
and computed including the 40\% scaling uncertainty in the nondiagonal 
covariance  matrix, results to be $10.93$, for $N=8$ data points, which is 
well within the 
$\sqrt{2N}$ range expected for a given subset of data in global fit and 
therefore should be considered as consistent. 
The value for $\chi^2$ is comparable to the one obtained ($11.2$) in a recent 
fit \cite{Hirai:2006sr} to both inclusive DIS data and the {\sc Phenix} 
measurement, although not including the scaling error 
in the computation. 

Similar agreement is found comparing the expectation of the fit 
for $\Delta g /g$ at $1$ GeV$^2$ against preliminary data from COMPASS
\cite{Ageev:2005pq,Procureur:2006sg}, and previous measurements 
\cite{Airapetian:1999ib,Adeva:2004dh}, as shown in Figure~\ref{deltag}b. 
In this case we include both the leading order (LO) and the NLO expectation 
because the reported values for $\Delta g /g$ correspond to a LO extraction 
improved with Montecarlo higher order corrections. 
We include in the plot the uncertainty band coming from a 2\% variation  in 
$\chi^2$, plus that coming from varying $Q^2$ up to $10$ GeV$^2$, what again
highlights the nice consistency between independent data set and the frameworks
implemented for the corresponding analyses. 

\section{Combined fit}.

\begin{figure}[hbt]
\includegraphics[width=4.5in]{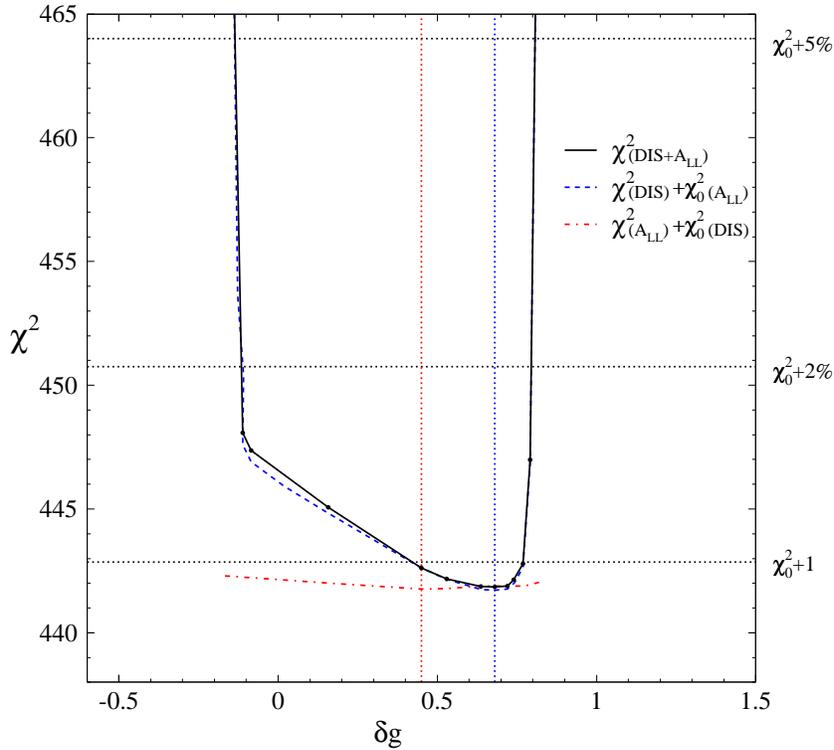}
\caption{Profiles of contributions to $\chi^2$ of different sets of data 
against the gluon net polarization. 
\label{profile}}
\end{figure}

Further insight on the interplay between DIS data and that coming from 
{\sc Phenix}
can be obtained analyzing the profile of $\chi^2$ function for the 
different subsets of data in a combined fit, against the range of variation of 
the net gluon polarization, as it was done in \cite{deFlorian:2005mw}. 
In Figure~\ref{profile} we show  the profile of the 
total $\chi^2${\scriptsize{$(DIS+A^{\pi^0}_{LL})$}} of a global fit to 
inclusive and semi-inclusive data along the lines of that of reference 
\cite{deFlorian:2005mw} but also including {\sc Phenix}  data as a solid line. 
The curve has similar shape to the one found in 
\cite{deFlorian:2005mw} but shifted upwards between eleven and thirteen units,
which is essentially the partial contribution of $A^{\pi^0}_{LL}$ to 
$\chi^2${\scriptsize{$(DIS+A^{\pi^0}_{LL})$}}.
In order to see the relevance of $A^{\pi^0}_{LL}$ data, this last contribution 
is also plotted as a dashed-dotted line with an offset 430.91 units, which the 
partial contribution  of DIS data  $\chi^2_0${\scriptsize$(DIS)$} for the best 
fit. 
Notice the partial contribution of $A^{\pi^0}_{LL}$ is almost flat around 
the minimum of $\chi^2${\scriptsize{$(DIS+A^{\pi^0}_{LL})$}} and reaches its 
own minimum for slightly 
lower values of $\delta g$ but within the one-unit variation of 
$\chi^2${\scriptsize{$(DIS+A^{\pi^0}_{LL})$}}, what highlights the 
consistency between both data sets.

\setlength{\unitlength}{1.mm}
In  Figure~\ref{profile} we have plotted also the profile obtained 
for DIS data only in \cite{deFlorian:2005mw} as a dashed line, 
with an offset given by the partial contribution of $A^{\pi^0}_{LL}$ 
at its minimun  $\chi^2_0${\scriptsize$(A^{\pi^0}_{LL})$}, in order to see the 
net effects on the fit of {\sc Phenix} data. In Table  \ref{tab:table1} we 
show the partial contributions to $\chi^2$ at both minima.

\begin{table}[hbt]
\begin{tabular}{ccccc} \hline \hline
  & $\chi ^2${\scriptsize$(DIS+A^{\pi^0}_{LL})$} &$\chi^2${\scriptsize$(DIS)$} & $\chi^2${\scriptsize$A^{\pi^0}_{LL})$}  & $\delta g$ \\ \hline
$\chi ^2_0${\scriptsize$(DIS+A^{\pi^0}_{LL})$}& 441.84&430.91&10.93&0.680 \\ 
$\chi ^2_0${\scriptsize$(A^{\pi^0}_{LL})$}& 442.63&431.82&10.81&0.450 \\ \hline \hline
\end{tabular}
\caption{\label{tab:table1} Partial contributions to $\chi^2$ values and 
first moment of $\Delta g$ at $Q^2=10$ GeV$^2$}
\end{table}

Clearly, these effects are almost imperceptible around the minimum but can 
be noticed for  $\chi^2${\scriptsize$(DIS+A^{\pi^0}_{LL})$} variations between 
the one-unit and the $2\%$ 
increase. Close to the minimum of 
$\chi^2${\scriptsize{$(DIS+A^{\pi^0}_{LL})$}}, the decrease of  
 $\chi^2${\scriptsize$(A^{\pi^0}_{LL})$} for decreasing $\delta g$ is 
overpowered by the increase of $\chi^2${\scriptsize$(DIS)$} and consequently,
the position of the minimum remains that found for DIS data.
Given the large number of DIS data included in the fit (478) compared to the
rather limited set of $A^{\pi^0}_{LL}$ available at present, the small impact
in the fit is not surprising, nevertheless the consistency shown, and the 
possibility of increasing considerably the statistics in the future is 
encouraging.

\section{Conclusions}
We have compared the constraints on the gluon polarization in the nucleon 
obtained in next to leading order global QCD fits to polarized deep inelastic 
scattering data with those coming  the double spin asymmetry measured by 
{\sc Phenix} at RHIC. 
Although the relative statistical weight of  $A^{\pi^0}_{LL}$ data in a 
global NLO fit including also DIS data is rather limited, we find a remarkable 
agreement and moderate improvement when combining both data sets. 
In the case of other direct measurements such as high-$p_T$ hadron production 
studied by COMPASS at CERN,  the lack of a NLO framework for the computation 
of the corresponding asymmetries does not allow to include them yet in a 
combined NLO global analysis however we find preliminary agreement which 
hopefully will be checked at NLO accuracy in the near future.

\section{Acknowledgements}

We warmly acknowledge Daniel de Florian for comments and suggestions 
and  Y. Fukao for discussions relative to {\sc Phenix} data.

\end{document}